\documentclass[conference]{IEEEtran}

\usepackage{amssymb}
\usepackage{amsmath}
\usepackage{booktabs}
\usepackage{caption}
\usepackage{subcaption}
\usepackage{hyperref}
\usepackage{tikz}
\usetikzlibrary{positioning,fit,calc,backgrounds,shapes.misc,arrows.meta,decorations.pathreplacing}

\definecolor{ipCwrap}{RGB}{222,235,247}   
\definecolor{ipRtl}{RGB}{252,233,217}     
\definecolor{ipHB}{RGB}{217,234,211}      
\definecolor{ipEdge}{RGB}{90,90,90}
\definecolor{ipDark}{RGB}{55,55,60}     
\definecolor{ipnameBg}{RGB}{240,240,240} 
\definecolor{ipChdr}{RGB}{42,98,150}    
\definecolor{ipRhdr}{RGB}{188,104,40}   

\usepackage{pgfplots}
\usepackage{xcolor}
\definecolor{todo}{RGB}{200,0,0}
\definecolor{todobg}{RGB}{255,255,150}

\usepackage{soul}
\usepackage{multirow}
\usepackage{listings}
\usepackage{tabularx}
\usepackage{comment}

\usepackage{graphicx}
\graphicspath{{}}
\DeclareGraphicsExtensions{.pdf,.jpeg,.png}

\begin{document}

\title{
ATLAS: Automated HLS for DL-Optimized FPGAs
}

\author{
\IEEEauthorblockN{Ruthwik Reddy Sunketa}
\IEEEauthorblockA{Arizona State University\\
Tempe, AZ, USA\\
rsunketa@asu.edu}
\and
\IEEEauthorblockN{Aman Arora}
\IEEEauthorblockA{Arizona State University\\
Tempe, AZ, USA\\
aman.kbm@asu.edu}
}

\maketitle

\begin{abstract}

FPGA architectures increasingly incorporate domain-specific in-fabric hardblocks to accelerate Deep Learning (DL) inference, particularly General Matrix Multiplication (GEMM), which dominates DL computation. To realize the performance gains of these hardblocks, manual RTL design is required: the programmer must understand the hardblock microarchitecture, instantiate them in RTL, and manage tiling and control logic. 
While programming in C/C++ and using HLS tools has increased the abstraction level and productivity of FPGA engineers, HLS tools do not support code generation for custom hardblocks natively.
Prior work has demonstrated that blackbox mechanisms in HLS tools can be used to target custom hardblocks, but this still requires explicit function calls in user-written HLS C and manual creation of RTL IP libraries, significant effort that must be repeated for every layer in a DL model. Furthermore, for DL, an even high-level programming interface, e.g., Pytorch/Keras instead of C/C++, is desirable for improved programmability and user adoption.

We present ATLAS, a fully automated flow from a high-level DL model description to a hardware implementation on an FPGA with custom in-fabric DL-optimized hardblocks, requiring no manual RTL design or explicit hardblock instantiation from the end user. Our approach uses GEMM as a universal abstraction layer and comprises two components: (1) \textbf{hls4ml-GEMM}, a compiler frontend that transforms DL layers into HLS C code with architecture-agnostic GEMM function calls, and (2) a \textbf{GEMM IP Generator}, an architecture-aware backend that produces hardblock-based RTL wrappers with tiling logic, control FSMs, and  scheduling metadata.
We evaluate the flow across 11 DL designs, including individual fully connected, convolution, and attention layers, as well as full CNN, MLP, and Transformer models targeting an FPGA architecture with Tensor Slices using Catapult for HLS and VTR for implementation.
Results demonstrate that, on the layer-wise designs, our automated flow reaches compute--area efficiency approaching that of a hand-written RTL baseline (approximately 89\%) while exceeding the soft-logic hls4ml baseline by 24\%. On the full designs, it attains a geomean efficiency of approximately 63\% of the RTL baseline and approximately 42\% above the soft-logic hls4ml baseline, reducing design time from days to hours.

\end{abstract}

\section{Introduction} \label{sec:intro}

\begin{figure}[!t]
    \centering
    \includegraphics[width=\linewidth]{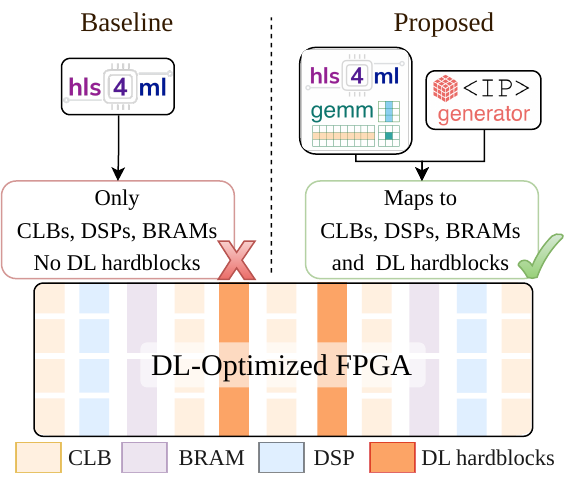}
    \caption{
    On a DL-optimized FPGA, the baseline hls4ml flow (left) maps DL workloads only onto soft-logic resources, leaving the specialized hardblocks unused. The proposed ATLAS flow (right) enables automatic mapping onto the DL hardblocks.
    }
    \label{fig:intro_overview}
\end{figure}

FPGA architectures increasingly include domain-specific in-fabric hardblocks to accelerate Deep Learning (DL) inference. These hardblocks typically target matrix multiplication, the computational bottleneck in DL workloads, and 
provide significant performance and energy efficiency improvements for DL applications over soft-logic\footnote{In this paper, we refer to CLBs, DSPs, and BRAMs as soft-logic, and specialized in-fabric DL compute blocks as hardblocks.} implementations. Examples include Intel AI Tensor Blocks~\cite{intel_ai_tensor_block}, Tensor Slices~\cite{arora:tensor_slice:TRETS}, and the Analog Dot Product Engines of AzureLily~\cite{azure_lily:2026}.

Programming these hardblocks requires RTL expertise. The end-user must understand the hardblock microarchitecture, tile computations to match its fixed dimensions, write Verilog designs with control FSMs, and manage data movement.
This effort must be repeated for every layer in a network, making end-to-end implementation of a full DL model labor-intensive and error-prone.

High-Level Synthesis (HLS) tools such as Vitis HLS~\cite{amd_vitis_hls}, Catapult~\cite{catapult_hls}, and Bambu~\cite{ferrandi2021bambu} improve design productivity by automating the translation of C/C++ into hardware. However, these tools only target soft-logic and lack native support for mapping computations onto custom architectural hardblocks, leaving these specialized resources underutilized.

Prior work~\cite{sunketa2026-RAW-hls} introduced a methodology using the HLS blackbox mechanism to target custom FPGA hardblocks from HLS. However, this approach has several limitations. First, the user must manually insert explicit function calls to target hardblocks into the HLS C source code. Second, the user must manually create the RTL IP that implements those function calls on the target hardblock.
Consequently, this approach does not scale to full, end-to-end models.

Existing FPGA deep-learning flows do not close this gap. Templated ML-to-FPGA frameworks (e.g., hls4ml~\cite{hls4ml-Schulte:2025mai}, FINN~\cite{FINN:umuroglu:fpt2017}) and MLIR-based compilers (e.g., SODA-OPT~\cite{soda:Agostini:2022}, ScaleHLS~\cite{scalehls:Ye:DAC21}) generate soft-logic implementations or synthesize new accelerators, but provide no general mechanism to target existing custom hardblocks. Vendor flows (e.g., Intel FPGA AI Suite) map to hardblocks but only on fixed commercial silicon, and cannot target novel FPGA architectures. 

We address this gap by developing an automated compilation flow, ATLAS, that translates a high-level DL model into a hardware implementation on a DL-optimized FPGA that maps relevant computations to DL hardblocks.
We use General Matrix Multiplication (GEMM) as an abstraction layer between our compilation flow and the DL hardblocks.
Our approach consists of two phases: first, an architecture-agnostic frontend converts a high-level DL network into an HLS C++ representation with embedded GEMM function calls to offload matrix multiplication computations in supported layers. 
Second, an architecture-aware backend accepts GEMM specifications from the frontend as input and generates an IP package containing RTL wrappers, tiling and control logic, and scheduling metadata that uses DL-optimized FPGA hardblocks.
This two-phase approach effectively isolates the computational requirements of a DL workload from the architectural properties of the target FPGA~(Figure~\ref{fig:intro_overview}).

In this paper, we make the following contributions:

\begin{itemize}
    \item To our knowledge, we propose the first architecture-agnostic, end-to-end automated flow that translates a high-level DL model description
    to a hardware implementation on an FPGA with custom in-fabric DL-optimized hardblocks, requiring no manual RTL design or explicit hardblock instantiation from the end user.
    \item We extend hls4ml to embed architecture-agnostic GEMM function calls within the generated HLS design for
    Dense, Convolution, and Attention 
    layers, offloading GEMM computations to the target hardblock architecture.
    \item We develop a GEMM IP Generator that accepts GEMM specifications (M, N, K, precision, and interface) and produces on-demand RTL wrappers with tiling logic, control FSMs, and HLS-compatible scheduling metadata.
    \item We evaluate the flow on common DL layers and full models, comparing both programmability
    and Quality-of-Results (QoR)
    against a hls4ml baseline and a hand-written RTL baseline, on an FPGA with Tensor Slices \cite{arora:tensor_slice:TRETS}, an academic DL hardblock.
\end{itemize}

We use Tensor Slices as a demonstration vehicle in this paper, but the flow generalizes to any in-fabric hardblock that performs a GEMM-like operation.

\section{Background and Related Work} \label{sec:related}

\subsection{GEMM and Domain-Specific Hardblocks}
\label{sec:related:gemm_hardblocks}

GEMM computes $\mathbf{C} = \mathbf{A} \times \mathbf{B}$, where $\mathbf{A}$, $\mathbf{B}$, and $\mathbf{C}$ are matrices of size is $M \times K$, $K \times N$, and $M \times N$ respectively. 
GEMM dominates computation in three major layer types in DL. Fully connected layers compute output as input multiplied by weights, which is a direct GEMM. Convolutions reduce to GEMM through the im2col transformation. Attention mechanisms contain both weight-stationary and dual-dynamic GEMMs: Q/K/V projections are weight-stationary matrix multiplications, while score computation $\mathbf{Q}\mathbf{K}^{T}$ and context computation $\text{softmax}(\dots)\mathbf{V}$ use runtime-generated operands.

DL-optimized FPGA hardblocks implement GEMM in hardware because it is typically the computational bottleneck in DL applications. 
Several domain-specific in-fabric hardblocks have been proposed for DL acceleration on FPGAs. Tensor Slices~\cite{arora:tensor_slice:TRETS} are GEMM processors integrated into the FPGA fabric. 
Each Tensor Slice accepts $8 \times 8$ INT8 activation and weight matrices, performs systolic multiplication, accumulates across the reduction dimension internally, to produce the result. 
It supports multiple precisions such as INT8, FP8, FP16, and BF16.
The Analog Dot Product Engines of AzureLily~\cite{azure_lily:2026} are compute-in-memory blocks that perform analog matrix-vector multiplication within RRAM crossbars. Intel AI Tensor Blocks~\cite{intel_ai_tensor_block}
have hardened dot-product units in Stratix NX devices. Newer Agilex devices integrate the same functionality into enhanced DSP slices.
All prior work on these hardblocks evaluates architectural benefits but assumes manual RTL design and does not address the programmability gap.

\subsection{HLS Tools and BlackBox Support}
\label{sec:related:hls}

Commercial HLS tools such as Catapult~\cite{catapult_hls} and Vitis HLS~\cite{amd_vitis_hls} provide a \emph{blackbox} mechanism that allows users to integrate external RTL blocks directly into an HLS-generated datapath. 
The user provides the RTL module definition, scheduling metadata (latency and initiation interval), and a tool-compatible interface specification; the HLS tool correctly schedules and integrates the block into the design, synthesizing only the remaining HLS logic to RTL while passing the blackbox RTL through unchanged.
Prior work used this mechanism to target Tensor Slices from HLS~\cite{sunketa2026-RAW-hls}, 
but required manually identifying which computations to offload, writing the RTL IP by hand, and annotating all scheduling constraints.

\begin{figure}
    \centering
    \includegraphics[width=\linewidth]{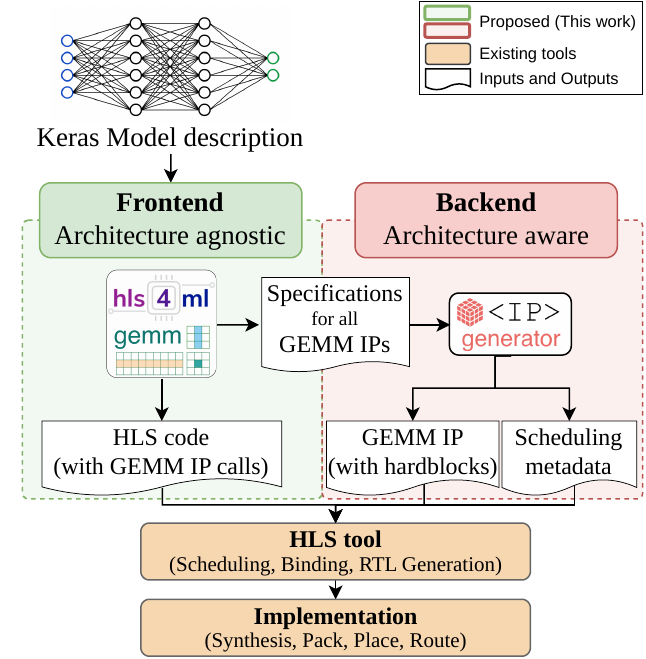}
    \caption{End-to-end ATLAS compilation flow. Architecture-agnostic hls4ml-GEMM frontend, combined with the architecture-aware IP generator, enables mapping a high-level Keras model onto a DL-optimized FPGA using GEMM as an abstraction.}
\label{fig:flow_overview}
\end{figure}

\subsection{Compilation Frameworks}
\label{sec:related:compilation}

DL-to-FPGA frameworks take high-level model descriptions and generate hardware implementations. 
hls4ml~\cite{hls4ml-Schulte:2025mai,hls4ml-Duarte:2018ite} and FINN~\cite{FINN:umuroglu:fpt2017} convert Keras and PyTorch models to HLS C using fixed templates and generate soft-logic implementations from LUTs, BRAMs, and DSPs, but neither targets custom hardblocks.
DNNWeaver~\cite{dnnweaver:fccm16} and AutoSA~\cite{autosa:fpga21} generate accelerator templates from high-level descriptions but do not map to existing hardblocks.

MLIR-based compilation flows leverage multi-level intermediate representations for hardware generation. SODA-OPT~\cite{soda:Agostini:2022} detects compute kernels in high-level code and synthesizes custom accelerators via HLS, generating new soft-logic hardware for each kernel rather than mapping to existing in-fabric DL-hardblocks. ScaleHLS~\cite{scalehls:Ye:DAC21,ye2022scalehls}, CIRCT~\cite{circt}, and Dynamatic~\cite{dynamatic:isfpga:2018} provide hardware compilation infrastructure but lack hardblock-aware lowering passes.

Vendor-specific DL compilation tools such as 
Intel FPGA AI Suite~\cite{intel_fpga_ai_suite} target vendor hardblocks through closed-source compilation. These tools are limited to existing commercial silicon and do not provide a general compiler mechanism applicable to academic or custom FPGA architectures.
Together, these approaches leave a gap between high-level DL model descriptions and custom in-fabric DL-optimized hardblocks. ATLAS closes this gap by using GEMM as a compiler-level abstraction between hls4ml-generated model code and architecture-aware hardblock IP generation, preserving an automated path through HLS and VTR without explicit hardblock calls or manually written RTL.
\section{Proposed Flow} \label{sec:proposal}

Figure~\ref{fig:flow_overview} illustrates the proposed ATLAS design flow. 
ATLAS provides an end-to-end, fully automated compilation path that accepts a high-level Keras model as input and produces a hardware implementation targeting an FPGA architecture with custom in-fabric DL-optimized hardblocks. 
ATLAS consists of two major components.  
\textbf{(1) hls4ml-GEMM (frontend):} A workload-aware, architecture-agnostic compiler pass that lowers the input model into HLS C code. For supported compute layers (Dense, Conv, Attention), it replaces standard templates with architecture-agnostic GEMM function calls while preserving the surrounding network dataflow and non-GEMM operations.
\textbf{ (2) GEMM IP Generator (backend):} An architecture-aware, workload-agnostic backend that maps each abstract GEMM instance onto the target hardblock by dynamically generating the necessary RTL wrappers, control logic, and scheduling metadata.

\subsection{The Interface between the Backend and the Frontend: GEMM Function Calls}
\label{sec:proposal:gemm_interface}

The boundary between the workload-aware frontend and the architecture-aware backend is strictly defined by a GEMM function call specification. Each GEMM instance is fully characterized by the parameters summarized in Table~\ref{tab:gemm_spec}.

The frontend specifies the dimensions and precision of all operands for every GEMM instance in the network. The \textit{Interface} parameter defines data delivery: in IO-stream mode, activations arrive sequentially via an AXI-Stream interface, whereas in IO-parallel mode, the activation matrix is exposed directly as an array to the GEMM IP.
The backend generator uses this parameter to instantiate the appropriate buffering and flow-control logic.
IO-stream uses the hls4ml streaming interface and its FIFO channels, with FIFO depths inherited from the baseline hls4ml flow, and IO-parallel exposes arrays as in baseline hls4ml. 
The boundary between the surrounding soft logic and the GEMM IP therefore follows the existing hls4ml interface convention rather than a new mechanism.
The \textit{Input mode} parameter defines the memory access pattern. Dense and Convolution layers utilize weight-stationary mode. Attention score ($Q\times K^{T}$) and context ($A\times V$) operations utilize dual-dynamic mode, as both operands are independent streams/arrays generated dynamically at runtime.

\begin{table}[!t]
\centering
\caption{Parameters of the GEMM function-call specification that defines the interface between the hls4ml-GEMM frontend and the IP generator backend.}
\label{tab:gemm_spec}
\small
\begin{tabular}{ll}
\toprule
\textbf{Parameter} & \textbf{Description} \\
\midrule
$M, N, K$ & GEMM dimensions \\
Precision & Operand data types from frontend \\
Interface & IO-stream or IO-parallel \\
Input mode & Weight-stationary or dual-dynamic \\
\bottomrule
\end{tabular}
\end{table}

This standardized specification effectively decouples the two components.
The frontend operates solely on the requirements of the layer, extracting the exact GEMM dimensions, data interface, and precision. The backend takes these parameters as input and abstracts away the physical hardware complexities; it automatically resolves tiling, padding, and scheduling to map arbitrary $M, N$, and $K$ dimensions onto the fixed geometry of the target hardblock.

\subsection{hls4ml-GEMM (Frontend)}
\label{sec:proposal:frontend}

The standard hls4ml framework~\cite{hls4ml-Schulte:2025mai} converts a high-level ML model into an internal graph of layer nodes, subsequently emitting parameterized C++ for each layer. These generated C++ designs define layer behavior using behavioral loop nests, localized arrays, and HLS pragmas. Downstream HLS tools synthesize this code into soft-logic datapaths, mapping them onto standard FPGA fabric resources (CLBs, DSPs, and BRAMs).
The baseline flow cannot directly exploit domain-specific hardblocks. 

We modify hls4ml at the graph and code-generation levels so that the generated HLS C code for supported layers replaces behavioral templates with the GEMM function calls defined in Section~\ref{sec:proposal:gemm_interface}. 
The frontend performs four transformations: weight pre-packing, dense mapping, convolution mapping, and attention mapping. 
These frontend transformations do not require any manual user annotations. The surrounding network architecture, including inter-layer dataflow and non-GEMM operations, remains unchanged from the baseline hls4ml flow.

\subsubsection{\textbf{Weight Pre-Packing}}

In the baseline hls4ml flow, weights are emitted in the layout expected by behavioral C++ templates, and the HLS tool generates runtime indexing and multiplexing logic to supply parallel values to the datapath. We introduce a compile-time weight pre-packing step that reorganizes the weight matrix into the exact access order required to match the stream or memory layout that GEMM IPs expect.
For convolutional layers, this reorganization also flattens the weight tensor to match the im2col layout used to lower the convolution to GEMM.
The pre-packed weights are emitted as a ROM array that can be accessed at an initiation interval of one cycle. This eliminates runtime transpose logic, reduces HLS-generated multiplexing, and ensures one weight transaction per cycle without stalls.

\subsubsection{\textbf{Dense Mapping}}

In the baseline hls4ml flow, Dense layers are implemented as matrix-vector or matrix-matrix computation using behavioral C++ templates. 
In stream mode, the generated code reads inputs into local arrays, invokes a dense compute kernel with HLS-inferred MAC trees, and writes outputs to the output stream. 
The \textit{latency strategy} aggressively unrolls the computation, while the \textit{resource strategy} time-multiplexes arithmetic based on the reuse factor. In both cases, the HLS tool is responsible for scheduling all arithmetic and memory access onto soft logic.

We replace Dense nodes with GEMM nodes. For a Dense layer with $K$ input features and $N$ output features, the computation is $\mathbf{A}[M \times K] \times \mathbf{B}[K \times N] = \mathbf{C}[M \times N]$, where $M$ is the batch, sequence, or frontend-selected tile dimension. 
Instead of the arithmetic loop body, the frontend emits two components: an activation row producer that provides input features into the GEMM IP, and a pre-packed weight ROM that provides weights. The GEMM IP computes the matrix product and produces an output stream.
This replacement eliminates the baseline's local compute arrays, MAC-tree inference, and large intermediate buffers. The HLS tool synthesizes only the surrounding interface logic to drive the GEMM IP inputs, and forwards results to the next layer.

\subsubsection{\textbf{Convolution Mapping}}

In the baseline hls4ml flow, convolution is implemented as a native, behavioral node using streaming window-generation structures. Conv1D uses shift-register windowing, while Conv2D uses line buffers to cache previous rows. Once a valid sliding window is formed, the generated C++ executes the multiply-accumulate (MAC) operations over the kernel volume, forcing the HLS tool to synthesize the corresponding arithmetic datapath in soft logic.

We preserve the streaming window-generation behavior but replace the compute portion with a GEMM. The frontend splits convolution into two components: a streaming im2col row producer and a GEMM node.
The im2col producer dynamically emits one $K$-wide activation row per valid output spatial position, where $K = H_k \times W_k \times C_{\text{in}}$ for a 2D convolution and $K =  W_k \times C_{\text{in}}$ for 1D convolution.
These flattened rows are directly streamed into the GEMM IP without materializing the full $[M, K]$ activation matrix in local memory.
The resulting GEMM operation is mapped with $M = H_{out} \times W_{out}$ (the total output spatial volume) and $N = C_{out}$ (the number of output filters). The corresponding filter weights are supplied from the pre-packed ROMs in a column-major format.

\subsubsection{\textbf{Attention Mapping}}

In the baseline hls4ml flow, Keras Attention layers are decomposed into lower-level Einsum and EinsumDense nodes representing $\mathbf{Q}$/$\mathbf{K}$/$\mathbf{V}$ projections, score computation ($\mathbf{Q} \times \mathbf{K}^T$), and context computation ($\text{scores} \times \mathbf{V}$). These nodes are emitted as behavioral C++ code, and the HLS tool synthesizes the corresponding tensor indexing, multiplication, accumulation, and transposition logic.

We intercept these Einsum nodes and map their core matrix operations to GEMM nodes.
The frontend parses each Einsum equation, identifies the free and contracting dimensions, and emits the corresponding GEMM specification. 
For example, to map the attention score computation, the frontend parses the Einsum equation:
\begin{equation}
    \mathcal{O}_{i,l_0,l_1} = \sum_{c=0}^{C-1} \mathcal{A}_{i,l_0,c} \mathcal{B}_{i,l_1,c}
\end{equation}
The compiler dynamically translates this tensor reduction into GEMM invocations:
\begin{equation}
    \forall i \in [0,I): \quad \mathbf{O}^{(i)} = \mathbf{A}^{(i)} \left(\mathbf{B}^{(i)}\right)^{T}
\end{equation}
where $\mathbf{A}^{(i)} \in \mathbb{R}^{L_0 \times C}$, $\mathbf{B}^{(i)} \in \mathbb{R}^{L_1 \times C}$, and $\mathbf{O}^{(i)} \in \mathbb{R}^{L_0 \times L_1}$.

EinsumDense nodes that implement $\mathbf{Q}$, $\mathbf{K}$, $\mathbf{V}$, and output projections map to weight-stationary GEMMs with pre-packed constant weights. Einsum nodes that implement $\mathbf{Q} \times \mathbf{K}^T$ and attention-value multiplication map to dual-dynamic GEMMs where both operands are runtime streams.
Non-GEMM operations, such as softmax, scaling, and normalization, remain in soft logic, similar to the baseline hls4ml.

\subsection{GEMM IP Generator (Backend)}
\label{sec:proposal:backend}

\begin{figure}[t]
    \centering
    \resizebox{0.9\columnwidth}{!}{%
    \begin{tikzpicture}[>=Latex,
        lyr/.style={draw=ipEdge, thick, rounded corners=3pt},
        func/.style={font=\small, text=black!75, align=left},
        hdr/.style={font=\normalsize, text=white!85, fill=ipDark, rounded corners=2pt, inner xsep=5pt, inner ysep=2.5pt},
        hb/.style={draw=ipEdge, thick, fill=ipHB, rounded corners=1.5pt,
                   minimum width=0.85cm, minimum height=0.3cm, font=\small}]
        \node[lyr, fill=ipCwrap, minimum width=8.6cm, minimum height=5.3cm] (cw) at (0,0) {};
        \node[hdr, fill=ipChdr] at (cw.north)
            {\textbf{C wrapper}\, \textcolor{white!75}{\small\itshape - HLS visible}};
        \node[func, anchor=north] at (0,2.2)
            {$\bullet$~Precision conversion\\ $\bullet$~FIFO buffering\\ $\bullet$~Blackbox orchestration};
        \node[lyr, fill=ipRtl, minimum width=5.0cm, minimum height=3.0cm] (rw) at (-1.45,-0.95) {};
        \node[hdr, fill=ipRhdr] at (rw.north)
            {\textbf{RTL wrapper}\, \textcolor{white!75}{\small\itshape - Blackbox}};
        \node[func, anchor=north] at (-1.45,0.3)
            {$\bullet$~Tiling\\ $\bullet$~Reduction\\ $\bullet$~Bias addition\\ $\bullet$~Control FSM};
        \foreach \x in {-2.95,-1.95,-0.95,0.05}{\node[hb] at (\x,-1.7) {HB};}
        \node[func, align=center, anchor=north] at (-1.45,-1.9) {Hardblock instantiations};
        \node[lyr, fill=ipDark!8, minimum width=2.4cm, minimum height=1.5cm, align=center, font=\small] (meta) at (2.75,-0.95)
            {\textbf{Scheduling}\\ \textbf{metadata}\\ \textcolor{black!65}{\scriptsize (latency, II)}};
        \draw[ipEdge, thick, dashed] (rw.east) -- (meta.west);
    \end{tikzpicture}}
    \caption{GEMM IP structure: The HLS-visible C wrapper handles precision conversion, FIFO buffering, and orchestration, exposing the RTL wrapper to the HLS tool as a blackbox annotated with its scheduling metadata,
    which implements tiling, reduction, bias addition, and a control FSM around multiple hardblock (HB) instances.}
    \label{fig:gemm_ip_structure}
\end{figure}
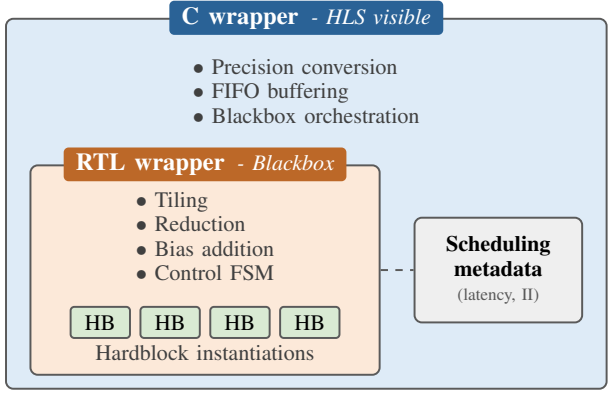

The GEMM IP Generator is the architecture-aware backend that accepts a GEMM specification ($M$, $N$, $K$, precision, input mode, interface) as input and produces an IP with a C++ interface that matches the function signature emitted by the hls4ml-GEMM frontend. 
This C++ function serves as a thin software shim around an RTL wrapper that instantiates the actual hardblocks. Using HLS blackbox mechanism, this C++ wrapper defines the port interface and scheduling metadata (latency and II) of the RTL wrapper while completely hiding its internal composition from the HLS tool.
Figure~\ref{fig:gemm_ip_structure} shows the generated IP structure with the HLS-visible C wrapper
and the blackbox RTL wrapper.

\subsubsection{\textbf{RTL Wrapper}}

Given a logical GEMM of dimensions $M \times N \times K$ and a hardblock that computes fixed-size $m \times n$ blocks with reduction-dimension depth $k$, the generator decomposes the computation into spatial tiles along the $M$ and $N$ dimensions (the generator instantiates $\lceil M/m \rceil \times \lceil N/n \rceil$ hardblock instances) and processes the $K$ dimension either spatially or temporally, depending on a user-selected strategy passed from the hls4ml configuration. 
Under the \textit{latency strategy}, the backend maximizes spatial parallelism by instantiating output-tile hardblocks across $M$ and $N$ and spatially unrolling $K$ where resources permit, accumulating partial products in a reduction tree. 
Under the \textit{resource strategy}, the backend reduces hardblock count by time-multiplexing tiles, reusing hardblock instances across the $K$ dimension and, in the general case, across subsets of the $M$ and $N$ tile grid.
When $K$ is processed temporally, the hardblock accumulates the partial products across $K$-steps internally; when $K$ is unrolled spatially, the wrapper combines the per-tile partial products in a reduction tree.
Reuse-factor-controlled time-multiplexing across the $M$ and $N$ output-tile grid is left as future work.
If $M$, $N$, or $K$ is not a multiple of the hardblock dimensions, the generator inserts zero-padding on inputs and masks invalid outputs.
By internally handling all the tiling logic, the generator hides the fixed geometry of the hardblock from the frontend. The frontend emits arbitrary GEMM dimensions, while the backend pads, tiles, and schedules the computation to match the hardblock geometry.

The generator also handles precision translation between the frontend's required precision and the hardblock's native precision. Hardblocks typically support a fixed set of operand precisions (e.g., INT8, INT16), whereas the frontend might require a diverse range of precisions.
hls4ml specifies the precision of each GEMM input and the required precision of the GEMM output. 
The IP generator selects a hardblock mode whose precision is equal to or wider than the precision specified by the frontend, 
handles conversion to 
precision supported by the hardblock, invokes the hardblock computation, and then converts the result back to the output precision required by hls4ml.
This fully decouples the frontend's precision requirements from the hardblock's fixed capabilities.
Any accumulation or bias addition performed outside the hardblock is carried out in a wider intermediate precision before the final output conversion.

\subsubsection{\textbf{Scheduling Integration}}

Commercial HLS tools like Catapult~\cite{catapult_hls} inherently treat blackboxes as standard C++ functions representing pure value transformations: they accept input arguments and return output values. Crucially, they do not support complex structs or high-level protocols (such as AXI-Stream) directly on blackbox boundaries.
This model conflicts with stateful hardblocks. For example, a Tensor Slice requires a cycle-accurate feed: it accepts activation and weight rows one beat per cycle, computes over a fixed multi-cycle latency (e.g., 16 cycles), and emits output rows one per cycle during a drain phase. If the C++
wrapper exposes the hardblock as a plain blackbox with a 16-cycle latency, the scheduler treats each input row as an independent, stateless operation. It can then insert stalls between rows or infer deep FIFOs for the data dependencies it perceives, which breaks the contiguous feed the hardblock's internal FSM requires.

To avoid this, the C wrapper as shown in Figure \ref{fig:gemm_ip_structure}
declares the blackbox as a single-cycle element (latency=1), invokes it inside a strictly pipelined ($II=1$) loop, and marks it as stateful (e.g., a \texttt{has\_state} pragma). The HLS tool therefore sees the blackbox as an ordinary function that takes real inputs and returns real outputs every cycle, and schedules exactly one call per cycle. The single-cycle latency keeps the scheduler from inserting deep pipeline buffers, and the stateful marking keeps it from reordering, merging, or stalling the calls. The C wrapper then manages three phases that the HLS tool does not see: feed, compute, and drain. The HLS tool treats every cycle as if it supplies an input and reads an output, but the loop asserts input-valid only during the feed phase and output-valid only during the drain phase; during the compute phase both the input and the output are dummy data. The blackbox RTL acts on these validity signals, so the true multi-cycle latency is absorbed by the loop while the HLS tool sees only a contiguous, single-cycles function call.
This workaround
enables integrating complex, multi-cycle hardblocks within standard HLS tools without compiler modifications.

\section{Methodology} \label{sec:method}

\subsection{FPGA Architecture}
\label{sec:method:arch}

The architecture consists of a conventional fabric of CLBs, DSPs, and BRAMs, augmented with columns of specialized Tensor Slice hardblocks. Each Tensor Slice is an $8 \times 8$ systolic array with a fixed configuration. Physical parameters, including Tensor Slice area (MWTA), delay, and latency, are derived from COFFE modeling at 22nm, as reported in~\cite{arora:tensor_slice:TRETS}. The VTR architecture file uses these parameters directly. 256 Tensor Slices are distributed in columns across the fabric.
The goal of this evaluation is not to compare Tensor Slices against other DL hardblock architectures, but to measure whether ATLAS can automatically target representative custom in-fabric DL hardblocks from a high-level model description.

\begin{table}[!t]
\centering
\caption{Benchmark suite: per-layer microbenchmarks and full end-to-end designs. Each microbenchmark isolates one layer type; each full design combines multiple layer types and exercises both weight-stationary and dual-dynamic GEMM modes. GEMM shapes $(M,N,K)$ are listed; convolutions use valid padding and unit stride.}
\label{tab:benchmarks}
\footnotesize
\setlength{\tabcolsep}{3.5pt}
\begin{tabular}{c l l l}
\toprule
& \textbf{Name} & \textbf{Layer config} & \begin{tabular}{c}\textbf{GEMM}\\\textbf{$(M,N,K)$}\end{tabular} \\
\midrule
\multirow{8}{*}{\rotatebox[origin=c]{90}{Layer-wise}}
& fc\_small & in $8{\times}8$, units 16 & $8{\times}16{\times}8$ \\
& fc\_large & in $24{\times}16$, units 16 & $24{\times}16{\times}16$ \\
\cmidrule(l){2-4}
& conv1d\_small & $L{=}33,C{=}4,K_s{=}2,F{=}16$ & $32{\times}16{\times}8$ \\
& conv1d\_large & $L{=}25,C{=}8,K_s{=}2,F{=}24$ & $24{\times}24{\times}16$ \\
\cmidrule(l){2-4}
& conv2d\_small & in $4{\times}6{\times}4$, $2{\times}2$, $F{=}7$ & $15{\times}7{\times}16$ \\
& conv2d\_large & in $4{\times}6{\times}6$, $2{\times}2$, $F{=}14$ & $15{\times}14{\times}24$ \\
\cmidrule(l){2-4}
& attn\_small & $S{=}8,D{=}8$, no proj & \begin{tabular}{l}$8{\times}8{\times}8$\\$8{\times}8{\times}8$\end{tabular} \\
& attn\_large & $S{=}16,D{=}16$, no proj & \begin{tabular}{l}$16{\times}16{\times}16$\\$16{\times}16{\times}16$\end{tabular} \\
\midrule
\multirow{3}{*}{\rotatebox[origin=c]{90}{Full designs}}
& MLP & 3 Dense & \begin{tabular}{l}$16{\times}16{\times}16$\\$16{\times}16{\times}16$\\$16{\times}8{\times}16$\end{tabular} \\
\cmidrule(l){2-4}
& CNN & \begin{tabular}{l}2 Conv\\Pool\\2 Dense\end{tabular} & \begin{tabular}{l}Conv: $64{\times}16{\times}32$\\Conv: $16{\times}32{\times}16$\\Dense: $16{\times}32{\times}32$\\Dense: $16{\times}16{\times}32$\end{tabular} \\
\cmidrule(l){2-4}
& Transformer & \begin{tabular}{l}MHA (4h)\\2 FFN\end{tabular} & \begin{tabular}{l}Q/K/V/O: $16{\times}16{\times}16$\\QK$^T$: $16{\times}16{\times}4$\\AV: $16{\times}4{\times}16$\\FFN1: $16{\times}32{\times}16$\\FFN2: $16{\times}16{\times}32$\end{tabular} \\
\bottomrule
\end{tabular}
\end{table}

\subsection{Tools}
\label{sec:method:tools}

We use Catapult 2026.1 for high-level synthesis and VTR 9 for logic synthesis, packing, placement, and routing. We choose VTR as our evaluation platform because it is the standard open-source framework for novel FPGA architecture research. Unlike commercial tools that are closed-source and tied to fixed silicon, VTR can model and evaluate custom DL-optimized architectures.
Catapult is an HLS tool that is not tied to a single backend and can target multiple hardware platforms, including both FPGAs and ASICs. Our choice of HLS tool is a matter of demonstration; the methodology can be easily extended to other HLS tools.

\subsection{Metrics}
\label{sec:method:metrics}

\subsubsection{Productivity}

We evaluate productivity using three metrics. User lines of code (LoC) measures the code written by the user: a Keras model and a hls4ml configuration file (precision, reuse factor, latency/resource strategy, and IO-stream/IO-parallel interface) for the hls4ml baseline and proposed flow, versus full Verilog for the RTL baseline. Generated LoC measures code produced by automation, including HLS C and RTL. Design time measures the estimated person-hours of manual effort, averaged across the benchmarks and assuming a graduate student already proficient with the relevant tools and methodologies.

\subsubsection{Quality of Results (QoR)}

We evaluate QoR using five metrics. Area is measured by resource counts (CLBs, DSPs, BRAMs, Tensor Slices), normalized to CLB-equivalents (CLB$=1$, DSP$=6.5$, BRAM$=3.12$, Tensor Slice$=28.5$) using per-resource Minimum Width Transistor Area (MWTA) ratios. Latency is the total cycles for one inference, reported from RTL simulation as the end-to-end latency from first input to last output. $F_{\max}$ is the maximum operating frequency reported by VTR. Throughput is computed as $\text{GMAC/s} = (\text{Total MACs} / \text{Latency}) \times F_{\max}$, with $F_{\max}$ in GHz. Compute--Area Efficiency is defined as throughput per unit area (GMAC/s per CLB-equivalent).

\begin{table}[!t]
\centering
\caption{Productivity comparison across the hls4ml baseline, hand-written RTL baseline, and proposed flow: user-written and generated lines of code, and estimated manual design time, averaged across the layer-wise and full designs.}
\label{tab:productivity}
\small
\begin{tabular}{lccc}
\toprule
\textbf{Flow} & \begin{tabular}{c}\textbf{User}\\\textbf{LoC}\end{tabular} & \begin{tabular}{c}\textbf{Generated}\\\textbf{LoC}\end{tabular} & \begin{tabular}{c}\textbf{Design}\\\textbf{Time}\end{tabular} \\
\midrule
hls4ml & 40 & 20{,}026 & $\sim$2--3 h \\
RTL & 1{,}885 & 0 & $\sim$7--8 d \\
\textbf{Proposed} & 40 & 13{,}941 & $\sim$2--3 h \\
\bottomrule
\end{tabular}
\end{table}

\subsection{Benchmarks}
\label{sec:method:benchmarks}

We evaluate the flow on multiple benchmarks organized into two categories. Microbenchmarks target individual layer types (fully connected, convolution, and attention).
Full models exercise end-to-end compilation across multiple layer types and GEMM modes (weight-stationary and dual-dynamic). Table~\ref{tab:benchmarks} lists both the microbenchmark and full-model configurations.
These benchmarks are selected to cover the supported layer types, the aligned and misaligned dimension cases, and both the weight-stationary and dual-dynamic GEMM modes, so that they exercise every frontend transformation and backend tiling path. Because the generator tiles any $(M, N, K)$ onto the fixed hardblock geometry, the mapping and control logic these benchmarks exercise is independent of the specific dimensions.

\subsection{Baselines for Comparison}
\label{sec:method:baselines}

We compare our proposed flow against two baselines. The first is standard hls4ml, which generates C++ from the Keras model; we then run this generated C++ through the HLS tool targeting soft logic. This baseline cannot utilize Tensor Slices.
The second is hand-written Verilog implementations of the same benchmarks that directly instantiate Tensor Slices. The RTL baseline is intentionally written to mirror the proposed flow's architecture and dataflow, including its hardblock mapping but without the HLS-generated interface logic, wrappers, and scheduling shims. This controlled baseline minimizes architectural differences between the manual and automated designs, allowing the comparison to isolate the overhead of automatic HLS-based generation relative to a manually optimized RTL implementation.
A given design can be implemented in many ways, using different dataflow, tiling, and architectural strategies. This evaluation is deliberately scoped to one such choice - the dataflow and architecture of the hls4ml's \textit{latency strategy}, which the RTL baseline reproduces by hand.
This comparison isolates the cost of HLS and automation for a fixed architecture and mapping. It is not a claim that the mapping itself is optimal. Evaluating automatically chosen mappings against an unconstrained, hand-optimized design is a separate question.

RTL simulation is used to verify functional correctness against the quantized Keras reference before QoR metrics are gathered. All three design flows are bit-exact to the quantized Keras model. 
\begin{table*}[!t]
\centering
\caption{QoR for layer-wise microbenchmarks and full designs. Per flow: CLB/DSP/TS are post-route counts; Fmx${=}F_{\max}$ (MHz), Lat${=}$latency (cyc), Tp${=}$throughput (GMAC/s), Ef${=}$compute-area efficiency (GMAC/s per 1000 CLB-eq; CLB${=}1$, DSP${=}6.5$, BRAM${=}3.12$, TS${=}28.5$). BRAM is $0$ everywhere and omitted. Geomean covers all eight layer-wise benchmarks, derived metrics only.}
\label{tab:qor}
\footnotesize
\setlength{\tabcolsep}{2.5pt}
\resizebox{\textwidth}{!}{%
\begin{tabular}{c l *{7}{c} | *{7}{c} | *{7}{c}}
\toprule
& & \multicolumn{7}{c}{hls4ml} & \multicolumn{7}{c}{\textbf{Proposed}} & \multicolumn{7}{c}{RTL} \\
\cmidrule(lr){3-9}\cmidrule(lr){10-16}\cmidrule(lr){17-23}
& \textbf{Benchmark} & CLB & DSP & TS & Fmx & Lat & Tp & Ef & CLB & DSP & TS & Fmx & Lat & Tp & Ef & CLB & DSP & TS & Fmx & Lat & Tp & Ef \\
\midrule
\multirow{9}{*}{\rotatebox[origin=c]{90}{Layer-wise}}
& fc\_small & 326 & 20 & 0 & 99.8 & 12 & 8.5 & 18.7 & 118 & 0 & 2 & 157.1 & 38 & 4.2 & 24.2 & 156 & 0 & 2 & 218.5 & 39 & 5.7 & 26.9 \\
& fc\_large & 625 & 40 & 0 & 95.9 & 29 & 20.3 & 23.0 & 274 & 0 & 12 & 157.0 & 62 & 15.6 & 25.3 & 288 & 0 & 12 & 163.8 & 63 & 16.0 & 25.4 \\
& conv1d\_small & 328 & 20 & 0 & 106.2 & 37 & 11.8 & 25.7 & 181 & 0 & 8 & 207.7 & 71 & 12.0 & 29.3 & 206 & 0 & 8 & 212.5 & 73 & 11.9 & 27.5 \\
& conv1d\_large & 900 & 57 & 0 & 94.9 & 30 & 29.2 & 22.9 & 349 & 0 & 18 & 168.6 & 71 & 21.9 & 25.4 & 370 & 0 & 18 & 181.4 & 73 & 22.9 & 25.9 \\
& conv2d\_small & 279 & 17 & 0 & 123.8 & 28 & 7.4 & 19.1 & 184 & 0 & 4 & 222.3 & 59 & 6.3 & 21.2 & 126 & 0 & 4 & 180.5 & 55 & 5.5 & 23.0 \\
& conv2d\_large & 770 & 53 & 0 & 83.0 & 28 & 14.9 & 13.4 & 323 & 0 & 12 & 165.5 & 74 & 11.3 & 17.0 & 256 & 0 & 12 & 166.0 & 69 & 12.1 & 20.3 \\
& attn\_small & 1538 & 36 & 0 & 61.6 & 299 & 0.2 & 0.1 & 1057 & 4 & 2 & 59.7 & 308 & 0.2 & 0.2 & 984 & 0 & 2 & 50.5 & 212 & 0.2 & 0.2 \\
& attn\_large & 4161 & 134 & 0 & 55.2 & 340 & 1.3 & 0.3 & 2700 & 6 & 16 & 55.1 & 349 & 1.3 & 0.4 & 1900 & 0 & 16 & 50.0 & 320 & 1.3 & 0.5 \\
\cmidrule(l){2-23}
& \textbf{Geomean} & -- & -- & -- & 87.3 & 49 & 6.1 & 6.1 & -- & -- & -- & 134.1 & 93 & 4.9 & 7.6 & -- & -- & -- & 134.1 & 87 & 5.2 & 8.6 \\
\midrule
\multirow{3}{*}{\rotatebox[origin=c]{90}{Full}}
& MLP & 1822 & 100 & 0 & 105.0 & 80 & 13.4 & 5.4 & 1160 & 0 & 20 & 143.0 & 130 & 11.3 & 6.5 & 1078 & 0 & 20 & 150.0 & 127 & 12.1 & 7.3 \\
& CNN & 10571 & 360 & 0 & 70.0 & 221 & 20.8 & 1.6 & 5389 & 0 & 128 & 95.0 & 357 & 17.4 & 1.9 & 4959 & 0 & 128 & 100.0 & 351 & 18.7 & 2.2 \\
& Transformer & 26261 & 766 & 0 & 54.6 & 1209 & 1.8 & 0.06 & 19822 & 13 & 89 & 60.0 & 898 & 2.7 & 0.12 & 7913 & 6 & 89 & 46.4 & 495 & 3.8 & 0.37 \\
\bottomrule
\end{tabular}%
}
\end{table*}

\section{Results} \label{sec:results}

\subsection{Productivity}
\label{sec:results:productivity}

Table~\ref{tab:productivity} compares lines-of-code written by the end user and estimated design time across the three design flows. The proposed flow requires the same user LoC as the hls4ml baseline, which consists only of a Keras model and configuration file, while achieving the same hardblock utilization as the RTL baseline. 
This eliminates thousands of lines of manual RTL. Days of manual effort are reduced to hours of compilation.

\begin{figure}[t]
    \centering
    \includegraphics[width=\linewidth]{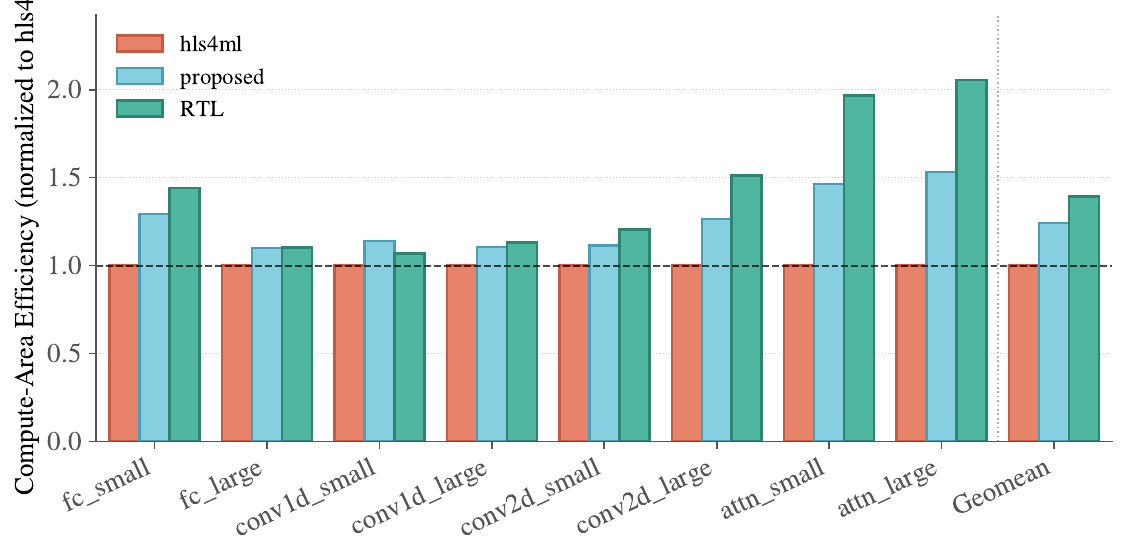}
    \caption{Compute--area efficiency (GMAC/s per 1000 CLB-equivalents) across microbenchmarks, normalized to the hls4ml baseline.
    The proposed flow improves efficiency by a geomean of 24\% by exploiting Tensor Slices, reaching approximately 89\% of the RTL baseline that requires manual RTL design.}
    \label{fig:efficiency}
\end{figure}

\subsection{Layer-wise QoR}
\label{sec:results:microbench}

We first evaluate QoR at the layer level. These microbenchmarks isolate the per-layer effect of the hls4ml-GEMM frontend mappings (Dense, Conv, and Attention), showing that each layer type is lowered onto Tensor Slices without per-layer manual effort before we compose them into full designs.

Table~\ref{tab:qor} reports the full per-layer QoR for all three flows (upper group). The three flows occupy distinct points in the resource, frequency, and latency space. On resources, the soft-logic hls4ml baseline maps every GEMM onto CLBs and DSPs, using the most soft logic and no Tensor Slices, whereas the RTL and proposed flows offload the matrix multiply onto Tensor Slices, removing almost all DSP usage and using far fewer CLBs; the proposed flow's resource usage is comparable to that of the RTL baseline.
The RTL and proposed flows reach a higher geomean $F_{\max}$ than the soft-logic baseline, with the proposed flow close to hand-written RTL. On raw latency and throughput the trend reverses: the soft-logic hls4ml baseline attains the lowest geomean latency and the highest geomean throughput, while the proposed flow again matches the RTL baseline. 
None of the evaluated designs consume block RAM. The streaming FIFOs and pre-packed weight ROMs are shallow and are implemented in flip-flops and LUTs rather than in dedicated block RAM.

By mapping GEMM onto Tensor Slices, the proposed flow attains a geomean compute--area efficiency 24\% higher than the hls4ml baseline, at the same user effort: a Keras model and config with no RTL coding. It reaches approximately 89\% of the RTL baseline's efficiency (geomean 7.6 vs.\ 8.6). These results validate that the proposed approach successfully maps a Keras model to an FPGA with Tensor Slices and utilizes them, reaching near-RTL efficiency while the automatically generated wrappers add only modest overhead, and requiring none of the hand-written RTL the baseline demands.

Crucially, the proposed flow delivers this efficiency at the productivity of the hls4ml baseline (Table~\ref{tab:productivity}): a Keras model and config, with no hand-written RTL. 
The efficiency itself originates from the Tensor Slice hardblock rather than from our flow; what the flow contributes is the ability to exploit it automatically, which previously required the manual RTL design of the baseline.

\subsection{Full-design QoR}
\label{sec:results:full_model}

\begin{figure}[t]
    \centering
    \includegraphics[width=0.8\linewidth]{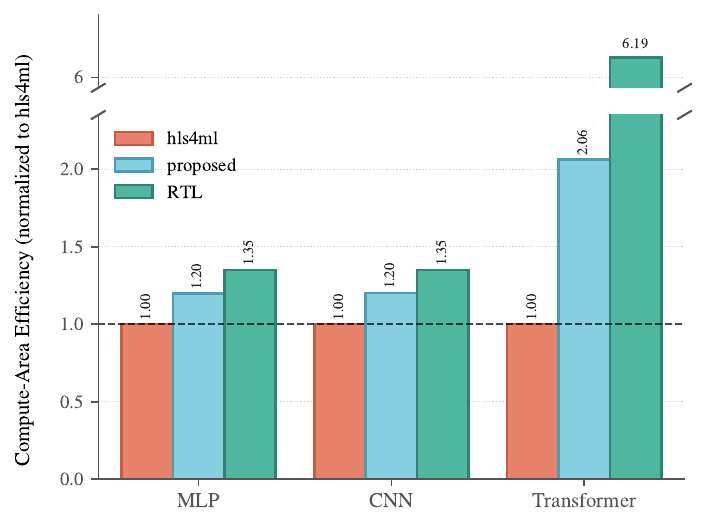}
    \caption{Compute--area efficiency (GMAC/s per 1000 CLB-equivalents) across full designs, normalized to the hls4ml baseline. The proposed flow exceeds the soft-logic baseline on all three models and approaches the RTL baseline on the dense-dominated MLP and CNN.}
    \label{fig:fullmodel_eff}
\end{figure}

Table~\ref{tab:qor} (lower group) and Figure~\ref{fig:fullmodel_eff} report end-to-end results for the full designs. On all three, the RTL and proposed flows offload the matrix multiply onto Tensor Slices, using fewer CLBs and DSPs than the soft-logic hls4ml baseline, and both improve on its compute--area efficiency. On the MLP and CNN designs, the proposed flow tracks the RTL baseline closely, reaching approximately 89\% and 86\% of its compute--area efficiency respectively, consistent with the layer-wise result, while the soft-logic baseline remains fastest on raw latency. 
For Transformer, dynamic attention GEMMs require additional soft-logic wrapper area, so the proposed automated flow uses more CLBs than the RTL baseline and reaches approximately 32\% of its compute--area efficiency on this design, though still roughly double that of the soft-logic hls4ml baseline.
Across the three full designs, the proposed flow attains a geomean compute--area efficiency of approximately 63\% of the RTL baseline and approximately 42\% above the soft-logic hls4ml baseline.
These results demonstrate that the proposed flow extends to complete networks that combine multiple layer types, retaining the efficiency advantage observed at the layer level while requiring no hand-written RTL. The Transformer model, which exercises both weight-stationary and dual-dynamic GEMM modes, demonstrates that the flow correctly handles the attention computation with dedicated GEMM IP instances.

\section{Discussion} \label{sec:discussion}

\subsection{The Cost of Automation}
\label{sec:discussion:cost}

Across the layer-wise benchmarks, the efficiency gap between the proposed flow and the RTL baseline is approximately 11\%, due to the resource overhead of standardized wrappers and FIFOs relative to hand-written RTL. This overhead arises from three sources: (1) stream interface FIFOs that decouple the HLS-generated datapath from the Tensor Slice, (2) padding logic for GEMM dimensions that are not multiples of the $8 \times 8$ hardblock geometry, and (3) control FSMs that are parameterized at generation time rather than hand-optimized per layer. We argue that this small overhead is an acceptable trade-off given the productivity gain from days of manual effort reduced to hours of compilation.

\subsection{The Generality of the GEMM Abstraction}
\label{sec:discussion:generality}

GEMM was chosen over lower-level primitives such as individual MACs or higher-level primitives such as full Conv2D blocks. GEMM is the sweet spot: it is high enough to encapsulate significant computation and amortize control overhead, yet low enough to serve as a universal building block for most DL layers. The im2col transformation maps convolutions to GEMM, and the Einsum decomposition maps attention to GEMM, both of which are well-established in the ML literature. GEMM is therefore the natural abstraction layer between a DL compiler and these hardblocks. It is simultaneously what the workload demands and what the hardware provides. This abstraction extends naturally to other existing and future hardblock designs that implement GEMM at different granularities.

Although we target Tensor Slices in this paper, this flow is agnostic to the specific hardblock architecture. Any hardblock that performs a GEMM-like operation can be targeted by updating the backend GEMM IP generator to handle its compute dimensions ($m, n, k$); the tiling, padding, control-FSM, and scheduling-metadata logic can be reused. 

\subsection{Applications and Usecases}
The main application of the proposed framework is to enable programming DL-optimized hardblocks that expose a GEMM or GEMM-like functionality. 
The idea is general though. When a new hardblock is introduced, the vendor CAD tools often lag the silicon by months or years.
The proposed flow offers a stop-gap path to programmability: by defining the new block's abstraction and adapting the backend IP generator to the new hardblock architecture, this flow can be used to drive the hardblock from a high-level model description without waiting for full CAD support. 
This lowers the barrier to adopting new FPGA features and shortens the time from silicon availability to usable acceleration.

An additional application of the proposed flow is rapid architecture exploration. Architecture researchers use open-source tools such as VTR~\cite{Elgammal:2025:TRETS:VTR-9} to propose and evaluate DL-optimized hardblocks, but lack a high-level software stack to program them, forcing architectural studies to rely on manual RTL that obscures the true productivity cost and system-level utility of the proposed hardware. By providing an automated path from a high-level model to a hardblock-based implementation, the proposed flow lets such studies evaluate new DL-optimized hardblocks without manual RTL design.

\subsection{Limitations}
\label{sec:discussion:limitations}

Currently, the flow performs a single open-loop compilation pass. The frontend selects the dataflow and per-layer GEMM configuration without feedback from the backend on the resulting latency, $F_{\max}$, or resource usage. As a result, the per-layer mapping cannot be tuned against measured hardware outcomes, and a globally balanced pipeline across layers is not guaranteed. 
This is a general characteristic of hls4ml, whose compilation
relies on user-specified directives such as the reuse factor rather than automated, feedback-driven optimization.

The current backend also does not yet fully exploit reuse-factor-controlled time-multiplexing across the complete $M$ and $N$ output-tile grid. The frontend presented in this paper fully supports this as a natural extension of the GEMM tiling model, but the backend implementation evaluated in this paper focuses on the tiling modes required by the benchmark suite. General reuse-factor-driven scheduling across $M$, $N$, and $K$ is not yet implemented.
The backend currently supports only configurations in which the hardblock's native precision is equal to or wider than the operand precision required by hls4ml; cases where the required precision exceeds the widest precision the hardblock supports are not yet handled.
Addressing these limitations is a direction for future work.
\section{Conclusion} \label{sec:conclusion}

We present ATLAS, a fully automated flow that uses GEMM as the abstraction layer between a high-level ML model and custom in-fabric DL-optimized hardblocks, pairing the hls4ml-GEMM frontend with an architecture-aware GEMM IP Generator backend to target Tensor Slices without manual RTL design. Evaluated across 11 DL designs spanning individual layers and full networks, ATLAS reaches a geomean compute--area efficiency of approximately 89\% of hand-written RTL (24\% above soft-logic hls4ml) at the layer level, and approximately 63\% of RTL (42\% above soft-logic hls4ml) on the full networks, while reducing design time from days to hours.

\bibliographystyle{IEEEtran}
\bibliography{refs}

\end{document}